\documentclass[12pt]{article}
\usepackage{graphicx,a4}

\def\th{\theta}

\def\la{\lambda}

\def\up{\upsilon}

\newcommand{\ben}{\begin{equation}}
\newcommand{\een}{\end{equation}}
\newcommand{\bea}{\begin{eqnarray}}
\newcommand{\eea}{\end{eqnarray}}
\newcommand{\ba}{\begin{array}}
\newcommand{\ea}{\end{array}}
\newcommand{\bi}{\begin{itemize}}
\newcommand{\ei}{\end{itemize}}

\def\math{\mathsurround 0pt}
\def\oversim#1#2{\lower.5pt\vbox{\baselineskip0pt \lineskip-.5pt
        \ialign{$\math#1\hfil##\hfil$\crcr#2\crcr{\scriptstyle\sim}\crcr}}}

\def\Re{\mathrm{Re}}
\def\Im{\mathrm{Im}}

\newcommand{\preprintno}[1]
{\vspace{-2cm}{\normalsize\begin{flushright}#1\end{flushright}}\vspace{1cm}}

\begin{document} 
\title{
\preprintno{SUSX-TH-98-023}
Sphalerons with CP-Violating Higgs Potentials}
\author{Jackie Grant\thanks{j.j.grant@sussex.ac.uk} and 
Mark Hindmarsh\thanks{m.b.hindmarsh@sussex.ac.uk}\\[3pt]
Centre for Theoretical Physics\\
University of Sussex\\
Brighton BN1 9QJ, U.K.}
\date{November 1998}

\maketitle
\vskip2pc

\begin{abstract}
We investigate the effect on the sphaleron in the two Higgs doublet 
electroweak theory of including CP violation in the Higgs potential. 
To have better control over the relation between the sphaleron energy and 
the physical quantities in the theory, we show how to 
parametrize the Higgs potential in terms of physical masses and mixing angles,
one of which causes CP violation. 
By altering this CP violating angle (and keeping the other physical quantities 
fixed) the sphaleron energy 
increases by up to 10\%.  We also calculate the static 
minimum energy path between adjacent vacua as a function of Chern-Simons number,
using the method of gradient flow.  The only effect CP violation has 
on the barrier is the change in height. As a by-product of our work on 
parametrization of the potential, 
we demonstrate that CP violation in the Higgs sector favours 
nearly degenerate light Higgs masses.  
\end{abstract}

\section{Introduction}
There has been much recent interest in the possibility of generating the 
  asymmetry in baryon
number observed today during an electroweak phase transition 
\cite{RubSha96,Tro98}. For a theory to allow any baryon 
asymmetry generation, it must satisfy certain conditions, originally identified by Sakharov \cite{Sak67}: 
a sufficient departure from thermal 
equilibrium, violation of charge conjugation (C) invariance and the combination 
of C with parity (P) invariance, and violation of baryon number conservation (B).

The standard model of Weinberg and Salam posseses all three properties. C is violated maximally in 
that only left handed fermions transform nontrivially under SU(2). CP is violated by a small amount given by the 
phase in the CKM matrix, and so CP is violated only through the Yukawa couplings.
B is conserved to all orders in perturbation theory at zero temperature, but can be violated non-perturbatively
either through the quantum tunneling of instantons between inequivalent vacua \cite{tHo76} or 
through the formation and decay of sphalerons \cite{KliMan84} at finite temperature \cite{KuzRubSha85}. 
The required departure from thermal equilibrium occurs during the sufficiently first order phase transition that 
breaks the gauge $\textup{SU}_L(2)\times \textup{U}_Y(1)\rightarrow \textup{U}_{em}(1)$.

Unfortunately, although the SM has all the ingredients for baryon asymmetry generation, it 
does not seem able to create 
the observed B asymmetry; the $CP$ violation is too small, and the phase transition is not sufficiently first order, 
and may even not be present at all (see for example \cite{Kaj+98} and 
references therein).

Attention has turned to extentions of the standard model. Of particular interest has been the two Higgs standard model, 
2HSM. This is a well motivated extension in that it allows sources of CP violation to enter through the Higgs 
potential, and it includes the Minimal Supersymmetric Standard Model as a subset.

It is the 2HSM model that we consider throughout. Curiously, although the sphaleron has been studied before in the 
2HSM \cite{PecKasZha91,BacTinTom96,Kle98}, the effect of 
CP violation in the Higgs potential has not been studied in detail. 
The purpose of this paper is to investigate how the sphaleron reacts to
CP violation in the Higgs sector.  In doing 
so we must decide how to parametrize CP violation, as 
this is by no means unique.  The most satisfactory solution, in our opinion, is the most physical one:  we parametrize 
CP violation by the mixing angle between the CP-odd and CP-even Higgs states (by CP-even and CP-odd we mean 
the quantum numbers the states would possess in the absence of mixing).  This leads us to the question of how 
we fix all 10 independent parameters in the most general SU(2)$\times$U(1)--invariant renormalisable two Higgs potential 
with acceptable levels of flavour-changing neutral currents. 
We solve this problem by inferring them from the Higgs masses and mixing angles, and the Higgs vacuum expectation value.
This leaves 2 parameters we must fix by hand, one CP violating, and the other 
a Higgs 4 point coupling.  This way of proceeding makes it much easier 
to satisfy experimental constraints and vacuum stability from the outset, 
although we must check that the resulting parameters 
also satisfy the constraints from boundedness of the potential.

Intriguingly, we find that it is quite difficult to cover a wide range of the CP-violating mixing angle, unless the 
CP-even states are almost degenerate in mass and the CP-odd state is very massive .  However, once we choose such a set of 
masses (100, 110, and 500 GeV), the sphaleron energy varies by as much as 10\% as the CP-odd mixing angle changes from 0 to about $\pi/3$.
This has important implications for the theory of electroweak baryogenesis. There are strong upper bounds on 
the lightest Higgs mass, which derive from the requirement that sphaleron transitions not wipe out any generated 
baryon asymmetry in the broken phase, which in turn constrains the strength of the transition.  
Even in the MSSM these bounds are close to being violated by the experimental lower bound on the Higgs mass 
\cite{HigMas}.

Another novelty of our approach is the use of the gradient flow method \cite{Man95} to find the entire
barrier, that is we calculate numerically the static minimum energy path (SMEP) between adjacent vacua as a function of 
the Chern-Simons number. This has not been done before for the two Higgs model, and we were motivated to do this 
to search for signs of the underlying CP asymmetry in the barrier.
The gradient flow technique is slightly easier to implement than methods which calculate the energy 
subject to a constraint \cite{AkiKikYan88,NolKun95}, and gives an intuitive feel for how the field 
might relax to the vacuum after a sphaleron has been created.

In Section 2, we introduce the model and our parameter definitions for the two Higgs potential.  We restrict our 
attention to SU(2), as we then need consider only spherically symmetric field configurations.  This is sufficient 
to demonstrate the point that one should consider more realistic Higgs potentials than hitherto.
In Section 3 we construct a general spherically symmetric ansatz for the sphaleron which allows
for CP-violation. In Section 4 we show how to determine the couplings of our potential as a function 
of the physical masses and 
mixing angles, and derive the bounds on these couplings from vacuum stability and boundedness of potential. We verify 
the sphaleron solution for a general 2HSM in Section 5 and proceed to find the entire barrier between neighbouring vacua 
by gradient flow. We then find the height of the energy barrier as a function of the only CP-violating physical 
parameter, $\theta$, 
the mixing between CP-even and CP-odd neutral Higgses. In Section  7 we supply our conclusions and outline possibilities
for future work.

\section{The Model}

We consider the bosonic sector of the electroweak lagrangian with two Higgs doublets in the limit $\theta_w=0$, 
so that
\ben
\label{e:lag}
{\cal L}=-\frac{1}{4}F^{a}_{\mu\nu}F^{a\mu\nu}
           +(D_\mu\phi_{\alpha})^\dagger (D^\mu\phi_{\alpha})
         -V(\phi_{1},\phi_{2})
\een
where           
\bea
F^{a}_{\mu\nu}&=&\partial_{\mu}W^{a}_{\nu}-\partial_{\nu}W^{a}_{\mu}
               +g\varepsilon^{abc}W^{b}_{\mu}W^{c}_{\nu}
\\       
D_{\mu}\phi_{\alpha}&=&(\partial_{\mu}+gW_{\mu})\phi_{\alpha}
\\
W_{\mu}&\equiv& W^{a}_{\mu}\frac{\sigma^{a}}{2i}
\eea
Higgs fields are labelled by $\alpha=1,2$.
The most general renormalizable potential for two Higgs doublets has 14 real couplings. To reduce the effect
of flavour changing neutral currents the symmetry $\phi_1\rightarrow-\phi_1$, $\phi_2\rightarrow\phi_2$ is often imposed but relaxed for 
dimension two terms \cite{gun}. Such a potential will have ten real couplings, and can be written
 \bea
V(\phi_{1},\phi_{2}) & = & (\lambda_{1}+\lambda_3)\left(\phi_{1}^{\dagger}\phi_{1}-\frac{\upsilon_{1}^2}{2}\right)^{2}
                        +(\lambda_{2}+\lambda_3)\left(\phi_{2}^{\dagger}\phi_{2}-\frac{\upsilon_{2}^2}{2}\right)^{2} \nonumber\\ 
                        & & +2\lambda_{3}\left(\phi_{1}^{\dagger}\phi_{1}-\frac{\upsilon_{1}^2}{2}\right)
                                   \left(\phi_{2}^{\dagger}\phi_{2}-\frac{\upsilon_{2}^2}{2}\right) \nonumber\\
                        & & +\lambda_{4}\left[\phi_{1}^{\dagger}\phi_{1}\phi_{2}^{\dagger}\phi_{2}
                                     -\Re^2( \phi_{1}^{\dagger}\phi_{2})      
                                     -\Im^2( \phi_{1}^{\dagger}\phi_{2})\right] \nonumber\\ 
& & +\lambda_{5}\left(\Re( \phi_{1}^{\dagger}\phi_{2})-\frac{\upsilon_{1}\upsilon_{2}}{2}\cos\xi\right)^{2}         
+\lambda_{6}\left(\Im( \phi_{1}^{\dagger}\phi_{2})-\frac{\upsilon_{1}\upsilon_{2}}{2}\sin\xi\right)^{2} \nonumber\\ 
& & +\lambda_{7}\left(\Re( \phi_{1}^{\dagger}\phi_{2})-\frac{\upsilon_{1}\upsilon_{2}}{2}\cos\xi\right)
\left(\Im( \phi_{1}^{\dagger}\phi_{2})-\frac{\upsilon_{1}\upsilon_{2}}{2}\sin\xi\right)       
\eea
the vacuum configuration is 
\ben
\phi_{\alpha}=\frac{\upsilon_{\alpha}}{\sqrt{2}}\left[\ba{c}0 \\ e^{i\theta_\alpha}\ea\right]
\een
where $\theta_1=0$ and $\theta_2=\xi$. We are free to redefine our fields, and we chose to write 
$\phi_{\alpha} \to \phi_{\alpha}e^{i\theta{\alpha}}$.
The potential can now be written as a function of nine real couplings
\bea
\label{e:pot}
V(\phi_{1},\phi_{2})& = &(\lambda_{1}+\lambda_{3})\left(\phi_{1}^{\dagger}\phi_{1}-\frac{\upsilon_{1}^2}{2}\right)^{2}
                        +(\lambda_{2}+\lambda_{3})\left(\phi_{2}^{\dagger}\phi_{2}-\frac{\upsilon_{2}^2}{2}\right)^{2} \nonumber\\
                        & &  +2\lambda_{3}\left(\phi_{1}^{\dagger}\phi_{1}-\frac{\upsilon_{1}^2}{2}\right)
                                   \left(\phi_{2}^{\dagger}\phi_{2}-\frac{\upsilon_{2}^2}{2}\right) \nonumber\\
                       & &  +\lambda_{4}\left[\phi_{1}^{\dagger}\phi_{1}\phi_{2}^{\dagger}\phi_{2} 
                                     -\Re^2( \phi_{1}^{\dagger}\phi_{2})      
                                     -\Im^2( \phi_{1}^{\dagger}\phi_{2})\right] \nonumber\\
   & &  +\lambda_{+}\left[\left(\Re( \phi_{1}^{\dagger}\phi_{2})-\frac{\upsilon_{1}\upsilon_{2}}{2}\right)^{2}
+\Im^2( \phi_{1}^{\dagger}\phi_{2})\right]\nonumber\\
  & & +\chi_{1}\left[\left(\Re( \phi_{1}^{\dagger}\phi_{2})-\frac{\upsilon_{1}\upsilon_{2}}{2}\right)^{2}
-\Im^2( \phi_{1}^{\dagger}\phi_{2})\right] \nonumber\\
  & &  +2\chi_{2}\left(\Re( \phi_{1}^{\dagger}\phi_{2})-\frac{\upsilon_{1}\upsilon_{2}}{2}\right)
\Im(\phi_{1}^{\dagger}\phi_{2})     
\eea 
where
\bea
\lambda_{+}&=&\frac{1}{2}(\lambda_{5}+\lambda_{6})\\
\lambda_{-}&=&\frac{1}{2}(\lambda_{5}-\lambda_{6})\\
\chi_{1}&=&\frac{\lambda_{7}}{2}\sin2\xi+\lambda_{-}\cos2\xi\\
\chi_{2}&=&\frac{\lambda_{7}}{2}\cos2\xi-\lambda_{-}\sin2\xi
\eea
The tenth coupling has been shifted in the redefinition to the CP transformation properties of the 
fields, so that now under a CP transformation 
$\phi_{\alpha}\rightarrow{-i}\sigma_{2}e^{-i2\theta_{\alpha}}\phi_{\alpha}^{\ast}$.

We chose Eq.\  \ref{e:pot} as our potential and set $\xi=0$ throughout, so that 
our vacuum configuration is invariant under CP, 
and the only source of  
CP violation is in the coupling $\chi_2$. The physical effect of $\chi_2$ will be to produce a non-zero 
mixing angle between the CP-even and CP-odd neutral Higgses.

\section{The Sphaleron Ansatz}

A  sphaleron is a static unstable finite 
energy field configuration, which represents the top of a static minimum energy path, SMEP, 
connecting inequivalent vacua. Each vacuum is 
distinguished in the unitary gauge by its Chern--Simons number ${n}_{\mathrm{CS}}$, defined as
\bea
{n}_{\mathrm{CS}} & =& \frac{g^2}{16\pi^2}\varepsilon_{ijk}\int d^3x\left[W^a_i\partial_{j}W^a_k
+\frac{1}{3}g\varepsilon^{abc}W^a_iW^b_jW^c_k\right] \\
&= &\frac{g^2}{32\pi^2}\int d^3xK^0,
\eea 
{where} $\partial_{\mu}K^{\mu}=F^a_{\mu\nu}\tilde{F}^{a\mu\nu}.$
The vacua have integer ${n}_{\mathrm{CS}}$, and are related by a large gauge transformation.

In the standard model there are twelve U(1) currents $J_u^{(i)}$, one for 
each left-handed fermionic species, defined as $J_{\mu}^{(i)}=\Psi_{L}^{(i)}\gamma^{\mu}\Psi_{L}^{(i)}$.
At the quantum level, these currents are not conserved:
\ben
\partial_{\mu}J^{\mu (i)}= \frac{g^2}{32\pi^2}F^a_{\mu\nu}\tilde{F}^{a\mu\nu}
\een
Hence the fermion numbers $N^{(i)}$ are not conserved, by an amount which by the change in the 
Chern-Simons number.  We can infer then that the baryon and 
lepton number violation as the background gauge fields change is 
\ben
\Delta \textup{B}=\Delta \textup{L}=3\Delta{n}_{\mathrm{CS}}
\een
Vacua with different Chern-Simons number are connected in configuration space along a path whose 
energy is always finite, so it is possible, with sufficient 
energy, to change B+L classically. The least energetic way of doing this is to form a sphaleron.

Motivated by \cite{PecKasZha91}, we chose a more general version of the spherically symmetric ansatz of 
\cite{RatYaf88}: 
\ben
\phi_{\alpha}=\frac{\upsilon_{\alpha}}{\sqrt{2}}(F_{\alpha}+{i}G_{\alpha}\hat{x}^a\sigma^{a})
\een
\ben
W_0=\frac{1}{g}A_0\hat{x}^a\frac{\sigma^a}{2i}
\een
\ben
W_i=\frac{1}{g}\left[\frac{(1+\beta)}{r}\varepsilon_{aim}\hat{x}_m
      +\frac{\alpha}{r}(\delta_{ai}-\hat{x}_a\hat{x}_i)
      +A_1\hat{x}_a\hat{x}_i\right]\frac{\sigma^a}{2i}
\een
where $F_{\alpha}=a_{\alpha}+ib_{\alpha}$ and $G_{\alpha}=c_{\alpha}+id_{\alpha}$.
      
On substituting our ansatz into Eq.\  \ref{e:lag}, our theory posseses spherical symmetry only if 
\ben
F_{\alpha}=\lambda(r,t) G_{\alpha}\label{const}
\een 
we assumed spherical symmetry, and verified that all terms that were zero because of Eq.\  \ref{const} did indeed vanish. 
With this condition the $\lambda_4$ term disappears from the potential. The $\lambda_4$ term is solely responsible for 
the mass of the physical charged Higgses, $m_{H^\pm}=\lambda_4{\upsilon^2}/{2}$. For our ansatz,  imposing 
spherical symmetry forces the $\la_4$ term in the potential to zero. Thus the charged Higgses decouple, 
and the choice of $\la_4$ is irrelevant for the sphaleron.

\section{Choosing Couplings}

We expanded 
\ben
\phi_{\alpha}=\frac{1}{\sqrt{2}}\left[\ba{c} H_{\alpha}^{+} \\ \upsilon_{\alpha}+H_{\alpha}^{0}\ea\right]
\een
to get the mass-squared matrix $M_G$ in the gauge basis $h_G$ of neutral scalar Higgses, 
where
\ben
h_G=\left[ \Re H^0_1, \Re H^0_2,A^0\right].
\een
Here, $A^0\equiv(\Im H^0_2\cos\beta -\Im H^0_1\sin\beta)$, and is odd under CP, the other two neutral Higgses being even. 
The 
components of the mass matrix in the gauge basis are
\bea
M_G(1,1)&=&\left[4(\lambda_1+\lambda_3)\cos^2\beta\label{e:m11}
	+(\lambda_++\chi_1)\sin^2\beta\right]\frac{\upsilon^2}{2}\\
M_G(1,2)&=&M_G(2,1)=(4\lambda_3+\lambda_++\chi_1)\cos\beta\sin\beta\frac{\upsilon^2}{2}\label{e:m12}\\
M_G(1,3)&=&M_G(3,1)=\chi_2\sin\beta\frac{\upsilon^2}{2}\label{e:m13}\\
M_G(2,2)&=&\left[4(\lambda_2+\lambda_3)\sin^2\beta\label{e:m22}
	+(\lambda_++\chi_1)\cos^2\beta\right]\frac{\upsilon^2}{2}\\
M_G(2,3)&=&M_G(3,2)=\chi_2\cos\beta\frac{\upsilon^2}{2}\label{e:m23}\\
M_G(3,3)&=&(\lambda_+-\chi_1)\frac{\upsilon^2}{2}\label{e:m33}
\eea
We denote the neutral Higgs fields and the mass-squared matrix in the physical basis by 
$h_P$ and $M_P$ respectively, with 
$
h_P=\left[h_2,h_1,h_3\right]
$and 
$
M_P={\mathrm{diag}}\left[m^2_2,m^2_1,m^2_3\right] 
$. The reason for our ordering convention is that in the absence of CP violation the lightest 
Higgs is conventionally the second entry in the physical higgs state vector $h_P$. 
We chose as our three mixing angles the Euler angles $\psi$, $\theta$, and $\phi$, such that
\bea
h_P&=&R_z(\psi)R_y(\theta)R_z(\phi)~h_G\equiv D(\psi,~\th,~\phi)~h_G\\
M_P&=&D(\psi,~\th,~\phi)~M_G~D^{-1}(\psi,~\th,~\phi)\label{e:mixing}
\eea
where 
\ben
R_z(\alpha)=\left[\ba{ccc}\cos\alpha & \sin\alpha & 0 \\
                           -\sin\alpha & \cos\alpha & 0 \\
                            0 & 0 & 1\ea \right],
~~~~~~R_y(\alpha)=\left[\ba{ccc}\cos\alpha &0 & -\sin\alpha \\
                           0 &1 & 0 \\
                            \sin\alpha & 0 & \cos\alpha\ea \right].
\een
We see that; $\phi$ is responsible for the mixing between the CP even Higgses to give $H^0$ 
and $h^0$, ($H^0=\Re H^0_1 \cos\phi +\Re H^0_2 \sin\phi$, and $h^0=-\Re H^0_1 \sin\phi +\Re H^0_2 \cos\phi$).
The angle $\theta$ is responsible entirely for the CP 
violation as it mixes the CP-odd $A^0$ with the 
CP-even $H^0$, and $\psi$ then shares this CP violation between $H^0$ and $h^0$.
For the case $\theta=0$, there is no CP violation; $m_1=m_{h^0}$, $m_2=m_{H^0}$, $m_3=m_{A^0}$, 
and the mixing angle between the two CP-even states is $\phi+\psi$.

If we invert Eq.\  \ref{e:mixing}, we can in principle find $M_G$ as a function of 
the physical parameters, which we call $\up^2X(m_1,m_2,m_3,\phi,\theta,\psi)$. 
The matrix $M_G$ is itself a function of eight unknown parameters ($\lambda_1,\lambda_2,
\lambda_3,\lambda_4,\lambda_+,\chi_1,\chi_2,\beta$) and one known parameter ($\upsilon=246$ GeV), where
\ben
\upsilon=\sqrt{\upsilon_1^2+\upsilon_2^2},~~~~~\tan\beta=\frac{\upsilon_2}{\upsilon_1}.
\een
The procedure is to find as many as posible of the parameters ($\lambda_1,\lambda_2,
\lambda_3,\lambda_4,\lambda_+,\chi_1,\newline \chi_2,\beta$) in terms of the physical quantities, which are 
four masses and three mixing angles ($m_1,m_2,m_3,m_{H^\pm},\phi,\theta,\psi$).  Clearly, we have to 
fix one of them by hand:  this we chose to be $\lambda_3$.

The charged Higgs mass $m_{H^{\pm}}$ is determined only by $\up$ and 
$\lambda_4$, and will only be relevant when we consider the boundedness of the potential and the stability 
of the vacuum.
We thus now have seven unknown couplings $\lambda_1,\lambda_2,\lambda_3,\chi_1,\chi_2,\beta$. From Eq.\  \ref{e:m13} 
and Eq.\  \ref{e:m23},
\bea
\chi_2 &=&2\sqrt{X(1,3)^2+X(2,3)^2},\label{e:chi2}\\
\beta &=&\arctan\left[\frac{X(1,3)}{X(2,3)}\right]
\eea
In the limit $\theta=0$; $X(1,3)=X(2,3)=0$, $\chi_2=0$, and $\beta$ can be taken as the limiting value of the ratio.
All other couplings are determined as below.

Using the non-independent relation ${\mathrm{Tr}}\; M_G=m_1^2+m_2^2+m_3^2$, and eliminating $\lambda_+$ using Eq.\  \ref{e:m33}, 
we were able to solve Eqs.\  \ref{e:m11}, \ref{e:m12} and \ref{e:m22} 
to determine all the couplings. On the understanding that $\lambda_3$ 
has already been fixed by hand, we get
\bea
\lambda_1&=&\left[\left(X(1,1)\cos\beta-X(1,2)\sin\beta\right)
              -2\lambda_3\cos2\beta\sin\beta\right]\frac{1}{2\cos^3\beta}\label{e:lambda1}\\
\lambda_2&=&\left[\left(X(2,2)\sin\beta-X(1,2)\cos\beta)\right)
              +2\lambda_3\cos2\beta\cos\beta\right]\frac{1}{2\sin^3\beta}\\
\lambda_+&=&-2\lambda_3+\left[X(3,3)+X(1,2)\frac{1}{\sin\beta\cos\beta}\right]\\
\chi_1&=&-2\lambda_3-\left[X(3,3)+X(1,2)\frac{1}{\sin\beta\cos\beta}\right]\label{e:chi1}
\eea
These are linear equations which are easily solved.

We now turn to the constaints that the parameters so derived must satisfy.  There are 
eight conditions on our potential, which derive from its boundedness and the stability of the vacuum state.
For boundedness of the potential we require that the eigenvalues of 
\ben
\frac{\partial^2 V(\phi_1,\phi_2)}{\partial x_i \partial x_j}
\een
are all positive, 
where $i,j=1,2,3,4$;  $x_1=\phi_1^{\dagger}\phi_1$, $x_2=\phi_2^{\dagger}\phi_2$, $x_3=\Re(\phi_1^{\dagger}\phi_2)$,
and $x_4=\Im(\phi_1^{\dagger}\phi_2)$. This gives the conditions
\bea
\lambda_1+\lambda_2+2\lambda_3&>&0\label{e:cond1}\\
4\lambda_1\lambda_2+4(\lambda_1+\lambda_2)\lambda_3&>&(4\lambda_3+\lambda_4)\lambda_4\\
\lambda_+&>&\lambda_4\\
(\lambda_+-\lambda_4)^2&>&\chi_1^2+\chi_2^2\label{e:cond4}
\eea
For a stable vacuum, we require 
\ben
m_1^2>0,~~~~~m_2^2>0,~~~~~m_3^2>0,~~~~~m_{H^{\pm}}>0.
\een
The last of these conditions is easily satisfied by $\lambda_4>0$.

On substituting Eqs.\  \ref{e:chi2}--\ref{e:chi1} into the inequalities \ref{e:cond1}--\ref{e:cond4} 
we could derive conditions directly on masses and mixing angles. In practice, 
we picked masses and mixings, calculated couplings for a suitable $\lambda_3$, and then verified 
that \ref{e:cond1}--\ref{e:cond4} held. Although we
 always used values of masses and mixing angles that satisfied all the conditions for 
boundedness and stability, in 
general it was quite difficult to find masses and mixings such that the inequalities \ref{e:cond1}--\ref{e:cond4} 
were satisfied. For example for $m_1$=100 GeV, $m_2$=300 GeV, $m_3$=400 GeV, $m_{H^{\pm}}$=50 GeV, 
$\psi$=$\phi$=$\frac{\pi}{8}$, the only allowed range of $\th$ is $2.40\leq\theta\leq2.85$.

\section{The Barrier}

We wanted to find the SMEP between vacua differing in ${n}_{\mathrm{cs}}$ by 1. 
This has been done in the one doublet case using 
the method of Lagrange multipliers to fix a constraint, either $n_{\mathrm{CS}}$ itself \cite{AkiKikYan88}, or 
the distance in field space from the sphaleron \cite{NolKun95}. We chose instead gradient flow \cite{Man95}. 
Gradient flow is defined from 
the static energy ${\cal E}_{\mathrm{s}}$ by
\ben
\frac{\partial f}{ \partial t} = -\kappa\frac{\delta {\cal E}_{\mathrm{s}}}{\delta f} 
\een
where $\kappa$ is a friction term, and $f$ are the fields of the theory. 
In our case $f\equiv(a_{\alpha},b_{\alpha},c_{\alpha},
d_{\alpha},\alpha,\beta, A_1)$ (note that in the static energy, $A_0=0$). Gradient flow will always describe 
a path of minimum energy since it forces the fields to evolve 
with a velocity orthogonal 
to the contours of ${\cal E}_{\mathrm{s}}$.

To flow down the barrier we first need to find the sphaleron. We chose to work in the radial gauge, $A_1=0$, which is
the most convenient, for in this 
gauge so fixing the boundary conditions of all fields fixes ${n}_{\mathrm{CS}}$. 
To start at the sphaleron, we fixed the boundary conditions corresponding to ${n}_{\mathrm{CS}}=\frac{1}{2}$, 
which are 
\bea
a_{\alpha}(0)&=&b_{\alpha}(0)=c_{\alpha}(0)=d_{\alpha}=\alpha(0)=0,~~~~~~~~~~\beta(0)=1,\\
a_{\alpha}(\infty)&=&b_{\alpha}(\infty)=d_{\alpha}(\infty)=\alpha(\infty)=0~~~~~~~~~~c_{\alpha}(\infty)=\beta(\infty)=1
\eea
We used Simultaneous Over Relaxation, and Chebyshev Acceleration (without even-odd ordering) \cite{C} to relax to the 
minimum energy configuration, 
and found a solution consistent  with \cite{PecKasZha91} 
for the same values of the parameters, in that we had the same energy and field profiles were indistinguishable 
when compared by eye. 
We used a grid size of 201, and considered our fields to have converged to solutions 
when the change in the absolute value of the fields integrated over the grid was  $<10^{-13}$.

This gave us the initial condition for our gradient flow. A technical problem now arises, because once 
the fields start flowing, $A_1$ has an equation of motion and will not in general remain zero, that is,
the configuration will not remain in the radial gauge.  It is useful to stay in the radial gauge in order
to be able to compute the Chern-Simons number easily. Hence, in order 
to keep in the radial gauge we carried out a gauge transformation
\bea
\tilde{a}_{\alpha}+i\tilde{c}_{\alpha}&=&e^{-i\frac{\Theta}{2}}(a_{\alpha}+ic_{\alpha})\\
\tilde{b}_{\alpha}+i\tilde{d}_{\alpha}&=&e^{-i\frac{\Theta}{2}}(b_{\alpha}+id_{\alpha})\\
\tilde{\alpha}+i\tilde{\beta}_{\alpha}&=&e^{-i\Theta}(\alpha+i\beta)\\
\tilde{A_1}&=&A_1-\Theta'
\eea
after each step of evolution of gradient flow, where we chose $\Theta$ so that
\ben
\Theta=\int A_1dx
\een
Naturally, there was no evolution unless we put a small perturbation in one of the fields. 
For a small positive perturbation we arrived at a vacuum with ${n}_{\mathrm{CS}}=1$, for a small negative perturbation we 
arrived at a vacuum with 
${n}_{\mathrm{CS}}=0$. The barrier was independent of choices of $\kappa$, and we used $\kappa=0.2$ throughout.

We checked our code by computing the barrier and the field profiles for the sphaleron in 
the one Higgs model, and comparing by eye the results of Nolte and Kunz \cite{NolKun95}, finding no noticeable 
differences.

\begin{figure}[ht]
\centering{
\includegraphics[height=8cm,width=9cm]{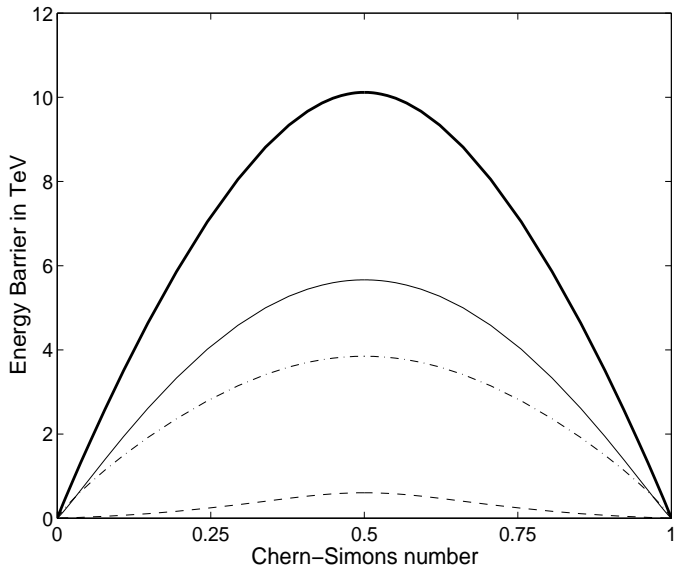}
}
\caption{\label{f:barrier} The sphaleron barrier for 
$m_1$=100 GeV, $m_2$=300 GeV, $m_3$=500 GeV, $\psi$=$\frac{\pi}{4}$, $\theta$=$\frac{\pi}{6}$, 
$\phi$=-$\frac{\pi}{4}$  and 
$\lambda_3$=0. The solid, dot-dashed and dashed lines are the contributions to the energy from the gauge field, the 
mixed or covariant derivative, and potential terms respectively.}
\end{figure}

The result, for one particular choice of masses and mixing angles, is shown in Fig.\ \ref{f:barrier}.
The barrier showed no unusual features for having an extra doublet, 
and CP violation.   In particular, it is symmetric around Chern-Simons number 
$\frac{1}{2}$.  That this should be so is not obvious {\it a priori}, because 
$n_{CS}$ is odd under CP, and  moreover changes by integer multiples
under gauge transformations.  Hence, a combination of a CP transformation 
and a large gauge transformation changes $n_{\mathrm{CS}}$ to 
$1 - n_{\mathrm{CS}}$.  It is therefore interesting that the barrier should 
be symmetric under this interchange when the underlying theory is not.  It is also 
interesting to note the small contribution of the potential energy.

\section{Energy as a function of CP violating angle}

\begin{figure}[ht]
\centering{
\includegraphics[height=8cm,width=9cm]{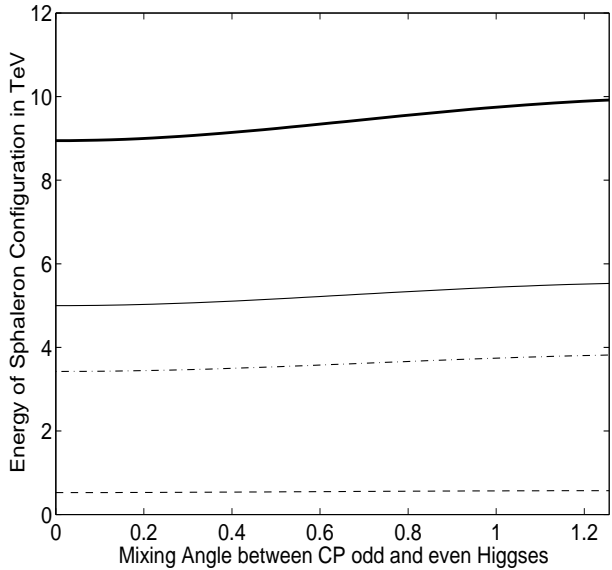}
}
\caption{\label{f:energy} The sphaleron energy as a function of the CP violating Higgs mixing angle 
$\th$, for Higgs masses $m_1$=100 GeV, $m_2$=110 GeV, $m_3$=500 GeV, and 
CP-conserving mixing angles $\psi=\frac{\pi}{8}$, $\phi=\frac{\pi}{8}$. The only coupling undetermined 
by the masses and the mixing angles is $\lambda_3$, which was set to zero.
The solid, dot-dashed and dashed lines are the contributions to the energy from the gauge field, the 
mixed or covariant derivative, and potential terms respectively.}
\end{figure}

We were unable to find sensible values of $m_1$, $m_2$, $m_3$, $m_{H^{\pm}}$, $\phi$ and $\psi$, that satisfied 
Eqs.\ \ref{e:cond1}--\ref{e:cond4} for the entire range 
$0\leq\theta\leq\pi$. In turned out that the largest ranges of $\theta$ favoured degenerate $m_1$ and $m_2$, large $m_3$, and small 
$m_{H^{\pm}}$. 
We plotted height of the energy barrier as a function of $\theta$ for the allowed range.  In Fig.\ 
\ref{f:energy} we display the energy as a function of the CP violating angle $\th$, for 
$m_1$=100 GeV, $m_2$=110 GeV, $m_3$=500 GeV, $\psi=\frac{\pi}{8}$, $\phi=\frac{\pi}{8}$  and $\lambda_3$=0, 
values which are consistent with current bounds on Higgs masses \cite{HigMas}.

We note that the energy changes from 9 to 10 TeV over the range of $\th$ we could obtain, or by roughly 10\%.  
This is not insignificant, in view of the tightness of the upper bound on the lightest Higgs mass, which comes 
from the requirement that the sphaleron rate be sufficiently low in the symmetry-broken phase at the bubble 
nucleation temperature $T_b$.  Increasing the 
sphaleron energy by this amount would mean that the ratio $\up(T_b)/T_b$ can be smaller, and hence that the 
transition can be more weakly first order.  This means that the lightest Higgs is allowed to be 
more massive (see, e.g., \cite{RubSha96} for a more detailed explanation).

Clearly, a more precise determination of the bounds must take into account finite temperature 
corrections, and use the true SU(2)$\times$U(1) theory. At the same time one should 
compute the ratio $\up(T_b)/T_b$ with the correct Higgs parameters. It is interesting to note that existing 
lattice calculations \cite{Kaj+98} assume that the Higgses in two-doublet models are not degenerate in mass, in order that 
all but the lightest can be integrated out.  We find that in order to get large range of 
CP-violating angles, one must have almost
degenerate light Higges.  It may well be worth recalculating the strength of the transition as a function 
of the (zero-temperature) Higgs masses in the almost degenerate case.

\section{Conclusions}

In this paper we have made a start on investigating the effect of 
CP violation in the Higgs potential on the sphaleron energy, by looking 
at the sphaleron in the pure SU(2) two Higgs theory.  We found firstly 
that the sphaleron energy changed by about 10\% as we scanned though a range of 
our CP-violating parameter, which we defined to be the mixing of the
CP-odd with the CP-even Higgs states.  This indicates that it is worth 
going on to 
calculate the sphaleron energy with finite temperature corrections taken into 
account, and in the full SU(2)$\times$U(1) 
two Higgs theory. In this way the bounds deriving from the requirement 
that the baryon asymmetry not be destroyed in the symmetry-broken phase 
can be more precisely worked out.  We also used a new method, that of gradient 
flow, to follow the fields as the sphaleron decays to the vacuum, 
and to exhibit the energy as a function of Chern-Simons number.
The shape of the barrier shows no qualitative difference from that in
CP-conserving theories: in particular, it is symmetric around Chern-Simons 
number $\frac{1}{2}$.  In a CP-violating theory, this feature is not {\it a priori} 
obvious, and perhaps points to a deeper symmetry in the equations.

Finally, in order to relate our sphaleron energies to physical quantities,
we adopted a novel strategy of computing as many as possible of the 
parameters of the Higgs potential from the masses and mixing angles of the 
Higgses.  With four masses, three mixing angles, and the 
Higgs vacuum expectation value we needed to choose only two of the parameters 
without physical input.  We found as a by-product that a large range of the 
CP-violating parameter could only be obtained if two of the Higgs were nearly 
degenerate in mass, and one much more massive.  This again has important 
implications for the lattice calculations of the strength of the electroweak phase 
transition, which hitherto have assumed that the Higgses are well separated in 
mass.

\section{Acknowledgements}
We are indebted to Paul Saffin for his programming help.  JG is supported by PPARC studentship 96314471 
and MH by PPARC grant GR/L56305.  Computational resources were made available by 
HEFCE and PPARC throught the JREI scheme.

\end{document}